\documentclass[10pt,conference,compsocconf]{IEEEtran}

\usepackage{times}

\usepackage{amsmath}
\usepackage{caption}
\usepackage{cite}
\usepackage[hyphens]{url}
\usepackage[hidelinks]{hyperref}
\usepackage{times}
\usepackage{xcolor}
\usepackage{tikz}
\usetikzlibrary{shapes,calc,arrows,decorations.markings}
\usepackage{graphicx}
\usepackage{orcidlink}
\usepackage{url}

\usepackage{breakurl}
\usepackage{hyperref}
\hypersetup{hidelinks,unicode,breaklinks}

\usepackage{booktabs}
\usepackage{threeparttable}

\usepackage{verbatim} 
\makeatletter
\renewcommand{\@biblabel}[1]{[#1]\hfill}
\makeatother
\usepackage{balance}

\usepackage{ifpdf}
\ifpdf
\pdfoutput=1 
\pdftrue
\fi

\makeatletter
\def\ps@IEEEtitlepagestyle{
	\def\@oddfoot{\mycopyrightnotice}
	\def\@evenfoot{}
}
\def\mycopyrightnotice{
	{\footnotesize
		\begin{minipage}{0.8\textwidth}
			\centering
			Please cite as: Simon Liebl, Leah Lathrop, Ulrich Raithel, Matthias Söllner, and Andreas Aßmuth, ``Threat Analysis of Industrial Internet of Things Devices,'' in \emph{Proc of the 11th International Conference on Cloud Computing, GRIDs, and Virtualization (Cloud~Computing~2020), Nice, France}, Apr 2020.
		\end{minipage}
	}
}
\makeatother
\makeatletter
\let\blx@rerun@biber\relax
\makeatother

\DeclareRobustCommand*{\IEEEauthorrefmark}[1]{%
	\raisebox{0pt}[0pt][0pt]{\textsuperscript{\footnotesize #1}}%
}

\tikzset{
	MyArrow/.style args={#1}{
		to path={let \p1 = ($(\tikztotarget)-(\tikztostart)$),
			\n1 = {int(mod(scalar(atan2(\y1,\x1))+360, 360))}, 
			\n2 = {veclen(\x1,\y1)} in \pgfextra{\typeout{\n1,\n2,\x1,\y1}} (\tikztotarget)
			node[
			#1 arrow, 
			#1 arrow head extend=1ex,
			draw,
			fill=gray!70,
			minimum height=\n2-\pgflinewidth-1,
			inner sep=0.7ex,
			rotate=\n1,%
			anchor=east,%
			]{}
	}},
	MyArrow/.default=single
}

\definecolor{blue1}{RGB}{30, 189, 247}

\begin{document}
	\pagenumbering{gobble}
	
	\title{\textbf{\Large Threat Analysis of Industrial Internet of Things Devices}}
	
	\author{%
		\IEEEauthorblockN{~\\[-0.4ex]\large Simon Liebl\IEEEauthorrefmark{*}\orcidlink{0000-0003-1311-4202}, Leah Lathrop\IEEEauthorrefmark{*}, Ulrich Raithel\IEEEauthorrefmark{$\dagger$}, Matthias Söllner\IEEEauthorrefmark{*} and Andreas Aßmuth\IEEEauthorrefmark{*}\orcidlink{0009-0002-2081-2455}\\[0.3ex]\normalsize}
		\IEEEauthorblockA{\IEEEauthorrefmark{*}Technical University of Applied Sciences OTH Amberg-Weiden, Amberg, Germany,\\Email: {\tt \{s.liebl $|$ l.lathrop $|$ m.soellner $|$ a.assmuth\}@oth-aw.de}}
		\IEEEauthorblockA{\IEEEauthorrefmark{$\dagger$}SIPOS Aktorik GmbH, Altdorf, Germany, Email: {\tt ulrich.raithel@sipos.de}\\[1ex]}
	}
	
	\maketitle
	
	\begin{abstract}
		As part of the Internet of Things, industrial devices are now also connected to cloud services. However, the connection to the Internet increases the risks for Industrial Control Systems. Therefore, a threat analysis is essential for these devices. In this paper, we examine Industrial Internet of Things devices, identify and rank different sources of threats and describe common threats and vulnerabilities. Finally, we recommend a procedure to carry out a threat analysis on these devices.
	\end{abstract}
	
	\renewcommand\IEEEkeywordsname{Keywords}
	
	\IEEEpeerreviewmaketitle
	
	\section{Introduction}
		
	Approximately 20 billion Internet of Things (IoT) devices are in use today \cite{gartner}, and this number could double in the next five years \cite{IDC}. The steadily increasing number of devices also raises the interest of attackers. During the first half of 2019, the overall number of cyberattacks increased by more than $350\%$ compared to the previous six months \cite{Fsec}. The majority of attacks either aim to infect IoT devices or to launch attacks using them, such as Distributed Denial of Service (DDoS) attacks. 
	
	The increasing number of attacks also affects Industrial Internet of Things (IIoT) devices. These are IoT devices specialized on industrial applications and used in Industrial Control Systems (ICSs) for holistic monitoring and analysis using cloud computing. A common approach is to integrate the IIoT functionality into existing low-power Operational Technology (OT) devices. This can be recognized by the number of OT devices connected to a network. While about $60\%$ of OT equipment was connected to the network in 2016, the figure had risen to almost $78\%$ by 2018 \cite{Fortinet}. 
	
	ICSs are a frequent target for attacks. Recently, Microsoft security researchers discovered that the hacker group APT33 focuses specifically on manufacturers, suppliers and maintainers of ICS components \cite{ICS-APT}. OT devices installed in an ICS can cause extensive damage, since they control physical processes. The impact can be severe, especially in critical infrastructures, where this can result in a breakdown of power or water supply, for example. The increasing number of OT devices connected to the network, however, increases the attack surface of ICSs. As a result, it becomes easier for hackers to attack, successfully exploit OT devices and cause damage to ICSs.
	
	Furthermore, the takeover of IIoT devices can also have an impact on cloud computing. In addition to the previously mentioned DDoS attacks on cloud servers, false data can be injected \cite{FalseData}. For example, ICS operators can be selectively supplied with incorrect information, e.g., abnormally high temperature values, to cause erroneous reactions, such as an emergency stop.
	
	As a consequence of the increasing threats, IIoT manufacturers must secure their devices to prevent such incidents. This requires awareness of the risks. It is important to understand who is interested in exploiting their device and what motivates attackers to do so. In this paper, we aim to identify the threats specific to IIoT devices, describe how attackers could proceed and support IIoT manufacturers in conducting a threat analysis for their devices. The paper is structured as follows: in Section~\ref{sec:IIoT}, the differences between IoT, IIoT and OT devices are clarified and the use of IIoT devices in ICSs are described. Different types of threat sources and their respective intentions are introduced in Section~\ref{sec:Sources}. In Section~\ref{sec:Threats}, several threats and vulnerabilities for IIoT devices are presented. A list of steps for a successful threat analysis follows in Section~\ref{sec:Procedure}. The paper concludes in Section~\ref{sec:Conclusion} with an outlook on further work.
	
	\section{The Industrial Internet of Things}
	\label{sec:IIoT}
	
	After a term differentiation, three potential setup options for a connection from IIoT devices to the cloud are described.
	
	\subsection{IoT, IIoT, OT and ICS}
	
	The IoT is a network of connected devices, which are sensors and/or actuators fulfilling a specific application \cite{IoTDefnTax}. Via the network they can, for instance, mutually exchange data or store and process data centrally and feed back the gained knowledge. This can be supported by cloud services. These have the advantage that there are already many semifinished solutions that simplify the integration of different devices. The number of devices or the required storage capacity can also be easily adapted, i.e., scalability. The use cases can be grouped in several categories, such as consumer applications (e.g., Smart Home), commercial (e.g., Medical and Healthcare) or infrastructure applications (e.g., Smart Grid). This paper focuses on industrial applications for which the already introduced term IIoT has been established. The main difference between IIoT and most IoT applications, such as consumer IoT, is that IoT services are human-centered and IIoT services are machine-oriented \cite{IIoTChallOpDir}. 
		
	The use of IIoT devices can have various advantages, such as boosting productivity, avoiding plant downtimes through predictive maintenance and reducing energy consumption. Furthermore, the IIoT should also enable products to be manu\-factured only after the order has been placed, i.e., build to order, and to be tracked by the customer during production and delivery. IIoT devices are usually part of the OT. OT can be found, for example, in industrial factories to monitor and control physical processes. The term was introduced to emphasize the significant difference to IT, such as field of application and used communication protocols. Some examples for OT/IIoT sensors are temperature probes or bar code scanners, actuators are, for instance, valves or power converter. The primary security challenges for IoT devices are privacy and confidentiality, e.g., human health data. However, IIoT devices focus additionally on safety and the impact on environment and society \cite{OTinICS}. They can potentially cause injury, death, damaged production equipment or environmental disasters. This can also affect large parts of the population through critical infrastructures, such as food or health.
	
	An ICS is usually structured into several layers. The lower levels are made up of OT devices and Programmable Logic Controllers (PLCs). The middle layers contain, for example, Human Machine Interfaces (HMIs) and engineering workstations. The top levels provide servers for services and backups. An increase in security can be achieved by dividing the ICS into multiple layers so that more protection can be provided to the lowest level, which is especially safety-critical. This concept is known as defense in depth. Another approach is air gapping, isolating the entire ICS network from the Internet or even corporate network. It has been demonstrated that particularly the latter does not provide sufficient security. Nevertheless, both measures result in more complex and expensive attacks. First, the IT network must be compromised (e.g., via email intrusion), then malware must be transferred to the OT network (e.g., via USB sticks) and, lastly, malicious code must be transferred to the PLCs \cite{ICSvec}. Once this is achieved, systems be controlled, damaged or spied on. However, these approaches conflict with the IIoT functionality of OT devices, as the lowest level requires Internet access. As a result, the architecture of ICS networks is affected by IIoT devices.
	
	\subsection{Cloud Connection Setups}
	
	Several IoT/IIoT architectures have already been proposed to implement segmented and logically structured networks \cite{Architectures}. In reality, however, these architectures can differ significantly. Therefore, different setups are only considered in an abstract way. The characteristics of a device, the task it performs and the level it is located on are important for the threat analysis.
	
	\begin{figure}[ht!]
		\centering%
		\footnotesize
		\begin{tikzpicture}[scale=0.99, every node/.style={scale=0.99}]
		\newcommand{\cloud}[3]{%
			\draw[thick, fill=#3, scale=0.9] (#1-1.6,#2-0.7) .. controls (#1-2.3,#2-1.1)
			and (#1-2.7,#2+0.3) .. (#1-1.7,#2+0.3) .. controls (#1-1.6,#2+0.7)
			and (#1-1.2,#2+0.9) .. (#1-0.8,#2+0.7) .. controls (#1-0.5,#2+1.5)
			and (#1+0.6,#2+1.3) .. (#1+0.7,#2+0.5) .. controls (#1+1.5,#2+0.4)
			and (#1+1.2,#2-1) .. (#1+0.4,#2-0.6) .. controls (#1+0.2,#2-1)
			and (#1-0.2,#2-1) .. (#1-0.5,#2-0.7) .. controls (#1-0.9,#2-1)
			and (#1-1.3,#2-1) .. cycle;
		}
		
		\def \wid {.4}
		\def \hei {.3}
		\def \tline {.15}
		\def \vheight {.1}
		\newcommand{\dev}[2]{%
			\draw[thick] (#1-\wid,#2-\hei) rectangle (#1+\wid,#2+\hei);
		}
		
		\newcommand{\plc}[2]{%
			\dev{#1}{#2}
			\draw[black, thick] (#1-\wid,#2-\tline) -- (#1+\wid,#2-\tline);
			\draw[black, thick] (#1-\wid,#2+\tline) -- (#1+\wid,#2+\tline);
			\draw[black, thick] (#1-\wid+.1,#2-\hei) -- (#1-\wid+.1,#2-\hei+\vheight);
			\draw[black, thick] (#1-\wid+.2,#2-\hei) -- (#1-\wid+.2,#2-\hei+\vheight);
			\draw[black, thick] (#1-\wid+.3,#2-\hei) -- (#1-\wid+.3,#2-\hei+\vheight);
			\draw[black, thick] (#1-\wid+.4,#2-\hei) -- (#1-\wid+.4,#2-\hei+\vheight);
			\draw[black, thick] (#1-\wid+.5,#2-\hei) -- (#1-\wid+.5,#2-\hei+\vheight);
			\draw[black, thick] (#1-\wid+.6,#2-\hei) -- (#1-\wid+.6,#2-\hei+\vheight);
			\draw[black, thick] (#1-\wid+.7,#2-\hei) -- (#1-\wid+.7,#2-\hei+\vheight);
			
			\draw[black, thick] (#1-\wid+.1,#2-.08) rectangle (#1-\wid+.5,#2+.08);
			
			\draw[black, fill=black] (#1-\wid+.075,#2+\hei-.1) rectangle (#1-\wid+.125,#2+\hei-.05);
			\draw[black, fill=black] (#1-\wid+.175,#2+\hei-.1) rectangle (#1-\wid+.225,#2+\hei-.05);
			\draw[black, fill=black] (#1-\wid+.275,#2+\hei-.1) rectangle (#1-\wid+.325,#2+\hei-.05);
			\draw[black, thick] (#1-\wid+.4,#2+\hei) -- (#1-\wid+.4,#2+\hei-0.1);
			\draw[black, thick] (#1-\wid+.5,#2+\hei) -- (#1-\wid+.5,#2+\hei-\vheight);
			\draw[black, thick] (#1-\wid+.6,#2+\hei) -- (#1-\wid+.6,#2+\hei-\vheight);
			\draw[black, thick] (#1-\wid+.7,#2+\hei) -- (#1-\wid+.7,#2+\hei-\vheight);
			
			\node[text width=1cm, text centered, scale=0.75] at (#1-\wid-0.25, #2) {PLC};
		}
		
		\newcommand{\edge}[2]{%
			\dev{#1}{#2}
			\draw[-latex] (#1,#2)--(#1,#2+0.25);
			\draw[-latex] (#1,#2)--(#1-0.25,#2-0.25);
			\draw[-latex] (#1,#2)--(#1,#2-0.25);
			\draw[-latex] (#1,#2)--(#1+0.25,#2-0.25);
			
			\node[text width=1cm, text centered, scale=0.75] at (#1-\wid-0.25, #2) {Edge};
		}
		
		\cloud{2.5}{.5}{black!5}
		\cloud{2.5}{1.75}{black!15}
		\cloud{-0.5}{.5}{black!10}
		\cloud{-0.5}{1.75}{black!20}
		\node[text width=2cm, text centered] at (-1, 1.6) {Plant Cloud};
		\node[text width=2cm, text centered] at (1.9, 1.6) {Device Manufacturer Cloud};
		\node[text width=2cm, text centered] at (-1, .3) {Company Cloud};
		\node[text width=2cm, text centered] at (1.9, .3) {Other Clouds};
		
		\draw (-2.25,-3.0) node[fill=blue1!40,minimum height=4cm, minimum width=2.75cm] (){};
		\draw (0.75,-3.8) node[fill=blue1!25,minimum height=2.5cm, minimum width=2.75cm] (){};
		\draw (3.75,-4.3) node[fill=blue1!10,minimum height=1.5cm, minimum width=2.75cm] (){};
		
		\node[inner sep=0pt] (scada) at (-3.15,-1.5) {\includegraphics[width=.8cm]{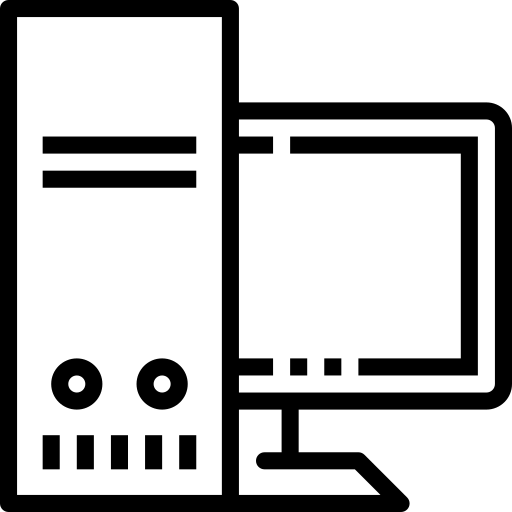}};
		\node[text width=1cm, text centered, scale=0.75] at (-2.35, -1.5) {SCADA};
		
		\draw[black, thick] (-3,-2.25) -- ++(0,0.38);
		\draw[black, thick] (-3.6, -2.25) -- ++(2.2,0);
		\draw[black, thick] (-1.75,-2.25) -- ++(0,-0.45);
		
		\plc{-1.75}{-3}
		\edge{0.75}{-3}
		
		\draw[black, thick] (-1.75,-3.75) -- ++(0,0.45);
		\draw[black, thick] (0.75,-3.75) -- ++(0,0.45);
		
		\draw[black, thick] (-3.6,-3.75) -- ++(2.7,0);
		\draw[black, thick] (-0.6,-3.75) -- ++(2.7,0);
		\draw[black, thick] (2.4,-3.75) -- ++(2.7,0);
		
		\draw[black, thick] (-3,-3.75) -- ++(0,-0.31);
		\draw[black, thick] (-1.5,-3.75) -- ++(0,-0.39);
		\draw[black, thick] (-0,-3.75) -- ++(0,-0.31);
		\draw[black, thick] (1.5,-3.75) -- ++(0,-0.39);
		\draw[black, thick] (3,-3.75) -- ++(0,-0.31);
		\draw[black, thick] (4.5,-3.75) -- ++(0,-0.39);
		
		\node[inner sep=0pt] at (-2.93,-4.45) {\includegraphics[height=.75cm]{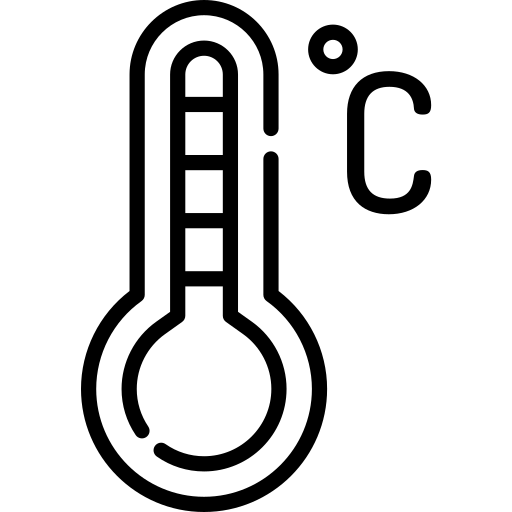}};
		\node[inner sep=0pt] at (-1.5,-4.5) {\includegraphics[height=.75cm]{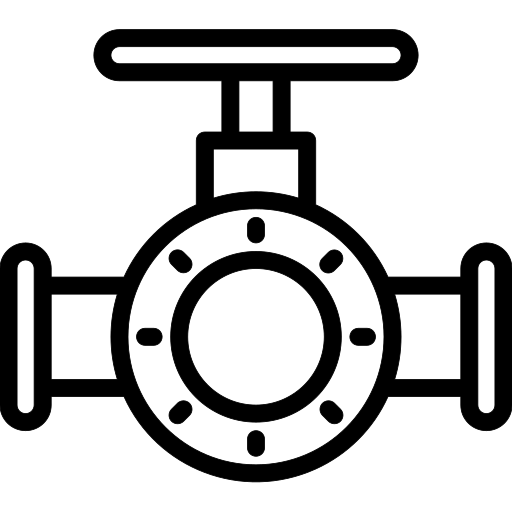}};
		
		\node[inner sep=0pt] at (.07,-4.45) {\includegraphics[height=.75cm]{thermometer.png}};
		\node[inner sep=0pt] at (1.5,-4.5) {\includegraphics[height=.75cm]{valve.png}};
		
		\node[inner sep=0pt] at (3.07,-4.45) {\includegraphics[height=.75cm]{thermometer.png}};
		\node[inner sep=0pt] at (4.5,-4.5) {\includegraphics[height=.75cm]{valve.png}};
		
		\draw (-2.25,-5.2) node[fill=blue1!45,draw,minimum height=.5cm, minimum width=0.6cm] (){\begin{minipage}{2.75cm}\begin{center}Large Setup\end{center}\end{minipage}};
		\draw (0.75,-5.2) node[fill=blue1!30,draw,minimum height=.5cm, minimum width=0.6cm] (){\begin{minipage}{2.75cm}\begin{center}Middle Setup\end{center}\end{minipage}};
		\draw (3.75,-5.2) node[fill=blue1!15,draw,minimum height=.5cm, minimum width=0.6cm] (){\begin{minipage}{2.75cm}\begin{center}Small Setup\end{center}\end{minipage}};
		
		\draw[] (-2,-2) edge[MyArrow=double] (-.5,-.75);
		\draw[] (0.75,-2.5) edge[MyArrow=double] (0.75,-.75);
		\draw[] (3.75,-3.5) edge[MyArrow=double] (2,-.75);
		
		\end{tikzpicture}
		\caption{Three possible setups for connections from IIoT devices to clouds.}
		\label{fig:Setups}
	\end{figure}
	
	Figure~\ref{fig:Setups} illustrates three possible setups. In small ones, each device can be connected separately to clouds. This could be, for example, a small, remote hydroelectric power plant connected to the Internet via mobile networks. The proprietary firmware of valves has been extended by a network stack for this purpose. The devices are connected to the operator's cloud for centralized monitoring and controlling and to the device manufacturer's cloud service for installing remote firmware updates.
	
	In the middle setup, devices are connected to the cloud via an edge gateway. It is not unusual for industrial devices to be older than ten years. They were not designed to send data to the cloud. Therefore, gateways collect data from several devices over mostly proprietary protocols, such as CAN or Modbus. Compared to low-power field devices, gateways have a more powerful processor and often a Linux-based operating system. 
	
	Even entire Supervisory Control And Data Acquisition (SCADA) systems are outsourced to the cloud in large industrial factories. Flexible web interfaces for desktops and mobile devices allow remote monitoring and control of the entire plant. In this scenario, many more connections to the cloud are possible, e.g., when the numerous field devices connect to their manufacturer's cloud or when all plants are combined in a company cloud.
	
	\section{Related Work}
	
	Since many IoT device manufacturers often prioritize functionality and time to market, security is neglected or not considered. This has been recognized by researchers and governmental institutions, leading to active research on the threats, necessary security requirements and mitigation techniques.
	
	The German Federal Office for Information Security (BSI) releases annually an Information Security Management System (ISMS), the so-called IT-Grundschutz Compendium, that covers, among others, technical and organizational aspects of information security \cite{BSI-IT}. The aspects are divided into several modules. For example, embedded devices (SYS.4.3), IoT devices (SYS.4.4) and ICS components (IND.2.1) are modules concerning threats and the resulting requirements.
	
	In \cite{IoTattacks}, Abomhara et al. evaluate IoT device attacks, vulnerabilities, assets and possible intruders. Although industrial systems, such as SCADA systems, are mentioned, the special characteristics of ICSs are not described in depth. In \cite{CIsecAnalysis}, Wurm et al. conducted a security analysis on a consumer IoT and an IIoT device and demonstrated how these devices could be exploited. However, the procedure is too specific and cannot be adapted to other devices.
	
	So far, manufacturers are assisted by standards and scientific papers in conducting a threat analysis for any system. However, there are no mandatory international guidelines on how the analysis should be carried out. In addition, computer-based threat modeling tools are not suitable for the special conditions of IIoT devices.  
	
	\section{Threat Sources and Motives}
	\label{sec:Sources}
	
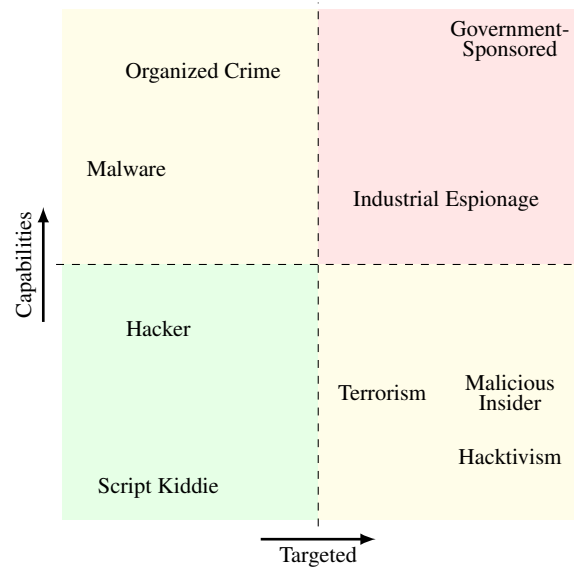
\begin{figure}[ht!]
		\centering%
		\footnotesize
		\begin{tikzpicture}[scale=0.85, every node/.style={scale=0.85}]
		
		\def \sqr {4cm}
		\draw (4,8) node[fill=yellow!10,minimum height=\sqr, minimum width=\sqr] (){};
		\draw (8,8) node[fill=red!10,minimum height=\sqr, minimum width=\sqr] (){};
		\draw (4,4) node[fill=green!10,minimum height=\sqr, minimum width=\sqr] (){};
		\draw (8,4) node[fill=yellow!10,minimum height=\sqr, minimum width=\sqr] (){};
		\draw[dashed] (1.9,6)--(10.1,6);
		\draw[dashed] (6,1.9)--(6,10.1);
		
		\node (v1) at (5,1.7) {};
		\node (v2) at (7,1.7) {};
		\draw[-latex, draw=black, line width=1pt] (v1) edge  node[below] {\normalsize Targeted}(v2);

		\node (v1) at (1.7,5) {};
		\node (v2) at (1.7,7) {};
		\draw[-latex, draw=black, line width=1pt] (v1) edge  node[above,rotate=90] {\normalsize Capabilities}(v2);
		
		\node[text width=3cm, align=center] at (9,9.5) 
		{\normalsize Government-Sponsored};
		\node[text width=3cm, align=center] at (4.2,9) 
		{\normalsize Organized Crime};
		\node[text width=3cm, align=center] at (3,7.5) 
		{\normalsize Malware};
		\node[text width=3cm, align=center] at (8,7) 
		{\normalsize Industrial Espionage};
		\node[text width=3cm, align=center] at (3.5,5) 
		{\normalsize Hacker};
		\node[text width=3cm, align=center] at (3.5,2.5) 
		{\normalsize Script Kiddie};
		\node[text width=3cm, align=center] at (7,4) 
		{\normalsize Terrorism};
		\node[text width=2cm, align=center] at (9,4) 
		{\normalsize Malicious Insider};
		\node[text width=3cm, align=center] at (9,3) 
		{\normalsize Hacktivism};
		
		
		\end{tikzpicture}
		\caption{Threat Sources.}
		\label{fig:Sources}
	\end{figure}

	To protect IoT devices from unauthorized access, it is helpful to know who is interested in using them, i.e., the threat sources. Depending on application and device characteristics, the sources can be different. For instance, IIoT applications in critical infrastructures are more likely to be attacked by Advanced Persistent Threat (APT) groups, whereas IoT devices with open Telnet or SSH ports are favored by botnet operators. Generally, there are also threats caused by natural disasters or unintentional misuse by employees, but these will not be considered in this paper. We have classified the sources based on two characteristics. First, to what extent the attack targets were selected arbitrarily or intentionally. Second, what capabilities attackers have, i.e., how many skills and financial resources are available to them. Figure~\ref{fig:Sources} classifies nine threat sources accordingly. In the following section, each source is described in detail.
	
	\subsection{Targeted attacks and capable attackers}	
	\paragraph{Government-Sponsored} The most serious threat arises when an ICS is the target of attackers who are supported by a government or agency. Examples include the attacks on the Iranian nuclear program (Stuxnet) \cite{Stuxnet} or on the Ukrainian power grid \cite{Ukraine}, both of which are suspected to have been supported by foreign governments. The attacks were targeted and only possible at high expense due to their complexity. The motives to conduct such attacks are usually political or economical.
	
	\paragraph{Industrial Espionage} Economic reasons are generally a major motive. Targeted attacks aim, for example, to sniff production figures, customer data and know-how, or simply cause financial loss to competitors. In recent years, there were several espionage attacks on German companies of the DAX (German stock index), including the ICS component manufacturer Siemens \cite{Espionage1}. 
	
	\subsection{Less targeted attacks, but capable attackers}
	\paragraph{Organized Crime} Organized cybercriminals try to blackmail their victims by encrypting sensitive data. The recently discovered ransomware EKANS seems to be specifically intended for ICSs because it terminates several common ICS-specific software processes \cite{OrgCrime}. 
	
	\paragraph{Malware} Malware is often designed to infect as many devices as possible, for instance, to build botnets. Mirai and its many variants demonstrated that millions of IoT devices are vulnerable to malware attacks \cite{TopBotnets}.
	
	\subsection{Targeted attacks, but less capable attackers}
	\paragraph{Terrorism} Threats from terrorism can be considered from two perspectives. There is a threat from extremist organizations. Although they are theoretically capable of carrying out attacks, few attacks are known in practice \cite{Extremists}. Additionally, terrorism can also be sponsored by states. Attacks on critical infrastructures, such as energy or water, affect the general civilian population. Therefore, they are a kind of terrorism. Since government-sponsored threats are already covered, the capabilities of terrorism is rated low.
	
	\paragraph{Malicious Insider} Insider attacks by (former) employees or contractors cause an average annual loss of more than eight million dollars \cite{InsiderCosts}. Employees, for example, could sell confidential data for personal financial gain or sabotage machines due to hostility towards the employer. They also possess specialist knowledge, which is particularly required for attacks on IIoT devices. Insider attacks are the major threat to OT \cite{InsiderOT}, especially for ICSs in critical infrastructures, as identified by an evaluation of US hydropower dams \cite{InsiderDams}.
	
	\paragraph{Hacktivism} The number of attacks by hacktivists is increasing and should therefore not be neglected. The attacks are targeted, but have not frequently been effective so far. Besides DoS attacks, attempts are made to steal data. This could affect, for instance, oil and gas companies or companies that make politically controversial decisions. The latter happened to heavy machinery maker Caterpillar Inc. as a result of the sale of bulldozers to Israel \cite{Hacktivsm}. 
	
	\subsection{Less targeted attacks and less capable attackers}
	
	\paragraph{Hacker and Script Kiddie} The last two threat sources we identified are hackers and script kiddies. The source code of malware, e.g., Mirai, is frequently published on code sharing platforms like Github or hacker forums. As a result, many people want to try them out for themselves. Compared to script kiddies, experienced hackers can build on this code and develop their own variants. 
	
	\section{Threats, Vulnerabilities and their Impact}
	\label{sec:Threats}
	
	Several threats were already mentioned in the listing of threat sources. In the following section, the threats are summarized briefly and common vulnerabilities are described. Possible attack vectors on IIoT devices are illustrated afterwards. Table~\ref{tab:ThreatsVuls} provides an overview of frequent threats and vulnerabilities for IIoT devices.
	
	\captionsetup{font={footnotesize,sc},justification=centering,labelsep=period}%
	\begin{table}[ht]
		\centering
		\caption{Common threats and vulnerabilities.}
		\begin{tabular}{p{3.4cm}p{3.4cm}}
			\toprule
			Threats & Vulnerabilities\\
			\midrule
			Abuse & Code execution \\
			Denial of Service & Communication manipulation \\
			Destruction & Design flaws and bugs \\
			Espionage & Memory manipulation \\
			Intellectual property theft & Misconfiguration \\
			Maloperation & Physical manipulation \\
			Ransomware & Privilege escalation \\
			Repudiation & Repudiation \\
			Spoofing & Web-based vulnerabilities \\
			\bottomrule
		\end{tabular}
		\vspace*{-0mm}
		\label{tab:ThreatsVuls}
	\end{table}
	\captionsetup{font={footnotesize,rm},justification=centering,labelsep=period}%
	
	\subsection{Threats}
	\paragraph{Abuse} The source of this threat could be malware or employees. The former utilizes IIoT devices as part of a botnet for DoS attacks, mining cryptocurrencies or for spreading spam. The latter could use the device for private purposes.
	
	\paragraph{Denial of Service} For ICS operators, the availability of all devices is most important because a single temporary breakdown can potentially lead to a production stop. Therefore, the failure of a device could have financial consequences for operators. A denial of service can be achieved not only by flooding devices with network requests but also by changing their configuration. Multiple devices could also be utilized to stop cloud servers. This would not only block one plant from its cloud services but all other plants of a large company.
	
	\paragraph{Destruction} The destruction of a device is also a form of denial of service, more precisely a permanent denial of service. The attack can be either on hardware or software. An example of the latter is BrickerBot, which destroyed more than ten million IoT devices \cite{BrickerBot}. Furthermore, the actuators of an OT device can be incorrectly triggered, destroying components, such as engines. The consequences are far more serious than a normal DoS attack. If there is no backup device that takes over immediately, the plant is out of operation. Additionally, data saved on the device may be lost. 
	
	\paragraph{Espionage} Espionage was already introduced in Section~\ref{sec:Sources}. Stealing production data, process procedures or even user data is often easy because many industrial communication protocols are not encrypted at all.
	
	\paragraph{Intellectual property theft} OT devices are usually specialized on one specific task. Manufacturers invest a lot of effort into their product in order to be better than competitors. As a result, leading manufacturers struggle with plagiarism and cloned, cheaply replicated hardware that runs their original firmware.
	\paragraph{Maloperation} Starting or stopping machines unexpectedly or making them work in slightly different ways is not a theoretical issue anymore.
	Two recent examples are TRITON \cite{Triton} and Industroyer \cite{Industroyer} that were specifically created for OT devices and protocols. The latter supports four industrial communication protocols and is capable of controlling switches and circuit breakers in electricity substations.
	
	\paragraph{Ransomware} If, in addition to the IT network, the OT network is also affected by a ransomware attack, some machines in the plant may no longer be available. As a result, the ICS must be shut down. This incident happened recently to a pipeline operator, who had to shut its operation down for two days, according to a report by the US Cybersecurity and Infrastructure Security Agency (CISA) \cite{Pipeline}. 
	
	\paragraph{Repudiation} In case of an error in an ICS, it should be possible to reconstruct the exact procedure with logs. Attackers could manipulate or delete them in order to remain undetected.
	
	\paragraph{Spoofing} IIoT devices must be uniquely identifiable. Attackers could masquerade as the device and send false data to PLCs or cloud services. The latest firmware could also be obtained by cloning original devices and spoofing their identity.
	
	\subsection{Vulnerabilities}
	
	\paragraph{Code execution} Arbitrary code execution is the goal of every attacker. Attacks can be either local or remote. Since the firmware of IIoT devices is mostly written in C/C++, they are vulnerable to memory attacks, such as buffer overflows.
	
	\paragraph{Communication manipulation} Message senders or receivers, measured values or commands can be easily manipulated due to unencrypted communication.
	
	\paragraph{Design flaws and bugs} Many industrial devices and protocols were not designed with security in mind. Even if this is the case, bugs can still occur. An example of this is the encrypted OPC UA protocol, which contained numerous flaws \cite{OPCUA}. This is particularly critical in ICSs because the firmware of the countless devices is rarely or never updated.
	
	\paragraph{Memory manipulation} By manipulating the memory, incorrect configurations can be loaded, faulty data can lead to inappropriate reactions and features that would be subject to additional costs can be unlocked illicitly.
	
	\paragraph{Misconfiguration} Misconfigurations enable many attacks. Common mistakes are unchanged default passwords, disabled firmware patches and open but unused ports.
	
	\paragraph{Physical manipulation} Attackers with physical access to IIoT devices can alter the hardware, e.g., sensors or actuators but also microcontrollers or memories. 
	
	\paragraph{Privilege escalation} Some actions should only be executed with higher privileges. For IIoT devices it is often simple to obtain these due to standard or company-wide passwords or backdoors of the developers. Furthermore, most industrial protocols do not support authentication. Therefore, it is not possible to verify authorization for them.
	
	\paragraph{Repudiation} The aforementioned threat is also a vulnerability, since insufficient logging and monitoring hinders the detection and verification of threats. Due to lack of identification mechanisms, actions can be easily repudiated.
	
	\paragraph{Web-based vulnerabilities} IIoT devices often run a web server for configuration, maintenance, monitoring or control of the devices. But this exposes them to web-based attacks. According to OWASP, the greatest risks include injection, broken authentication and cross-site scripting (XSS) among others \cite{OWASP}.
	
	\subsection{Attack Vectors}
	
\begin{figure}[t!]
		\centering%
		\footnotesize
		\begin{tikzpicture}[scale=0.80, every node/.style={scale=0.80}]
		
		\newcommand{\interface}[2]{%
			\def \L {-0.12}
			\def \R {-\L}
			\def \T {1.2}
			\def \B {-\T}
			
			\def \AD {0.15} 
			\def \APD {0.2} 
			\def \AIH {0.2} 
			\def \AIM {0.1} 
			\def \AOH {0.15} 
			\def \ALX {\L-\AD}
			\def \ALPX {\ALX-\APD}
			\def \ALYS {-0.65}
			\def \ALYE {\ALYS-\AIH}
			\def \ALYM {\ALYS-\AIM}
			
			\def \ARX {\R+\AD}
			\def \ARPX {\ARX+\APD}
			\def \ARYS {-\ALYS}
			\def \ARYE {\ARYS+\AIH}
			\def \ARYM {\ARYS+\AIM}
			\draw[thick] (#1+\L,#2+\B) -- (#1+\L,#2+\ALYE);
			\draw[thick] (#1+\L,#2+\ALYS) -- (#1+\L,#2+\T);
			\draw[thick] (#1+\ALX,#2+\ALYE) -- (#1+\L,#2+\ALYE);
			\draw[thick] (#1+\ALX,#2+\ALYE) -- (#1+\ALX,#2+\ALYE-\AOH);
			\draw[thick] (#1+\ALX,#2+\ALYS) -- (#1+\L,#2+\ALYS);
			\draw[thick] (#1+\ALX,#2+\ALYS) -- (#1+\ALX,#2+\ALYS+\AOH);
			\draw[thick] (#1+\ALX,#2+\ALYS+\AOH) -- (#1+\ALPX,#2+\ALYM);
			\draw[thick] (#1+\ALX,#2+\ALYE-\AOH) -- (#1+\ALPX,#2+\ALYM);
			\draw[thick] (#1+\L,#2+\T) -- (#1+\R,#2+\T);
			\draw[thick] (#1+\R,#2+\ARYE) -- (#1+\R,#2+\T);
			\draw[thick] (#1+\R,#2+\B) -- (#1+\R,#2+\ARYS);
			\draw[thick] (#1+\ARX,#2+\ARYS) -- (#1+\R,#2+\ARYS);
			\draw[thick] (#1+\ARX,#2+\ARYS) -- (#1+\ARX,#2+\ARYS-\AOH);
			\draw[thick] (#1+\ARX,#2+\ARYE) -- (#1+\R,#2+\ARYE);
			\draw[thick] (#1+\ARX,#2+\ARYE) -- (#1+\ARX,#2+\ARYE+\AOH);
			\draw[thick] (#1+\ARX,#2+\ARYE+\AOH) -- (#1+\ARPX,#2+\ARYM);
			\draw[thick] (#1+\ARX,#2+\ARYS-\AOH) -- (#1+\ARPX,#2+\ARYM);
			\draw[thick] (#1+\L,#2+\B) -- (#1+\R,#2+\B);
		}
		
		\def \cd {gray!5} 
		\def \cdt {gray!10}
		\def \ca {gray!15} 
		\def \cat {gray!20}
		\def \clic {blue1!10}
		\def \clict {blue1!15}
		\def \cpc {blue1!25}
		\def \cpct {blue1!30}
		\def \cma {blue1!40}
		\def \cmat {blue1!45}
		
		
		\draw (0.25,-2) node[fill=\cd,minimum height=3cm, minimum width=9.4cm] (){};
		\node[draw, rotate=90, minimum width=2.45cm, minimum height=1.1cm,fill=\cdt] at (-5,-2.25) {\normalsize Hardware};	
		\node[text width=3cm, align=left] at (-3.9,-1.25) {\textbf{\footnotesize Zone 0}};
		
		\draw (0.25,0.25) node[fill=\ca,minimum height=2.5cm, minimum width=9.4cm] (){};
		\node[draw, rotate=90, minimum width=2.45cm, minimum height=1.1cm,fill=\cat] at (-5,0.25) {\normalsize Software};	
		\node[text width=3cm, align=left] at (-3.9,1.25) {\textbf{\footnotesize Zone 1}};
		
		\node[text width=3cm, align=center] at (0,-1) {\normalsize Device};
		\draw[thick, rounded corners] (-0.6,-1.6) rectangle +(1.2,1.2);
		\draw[thick] (-1.2,0) circle (.075);
		\draw[thick] (-1.2,-1) circle (.075);
		\draw[thick] (-1.2,-2) circle (.075);
		\draw[thick] (1.2,-0) circle (.075);
		\draw[thick] (1.2,-1) circle (.075);
		\draw[thick] (1.2,-2) circle (.075);
		\draw[thick] (0,0) circle (.075);
		\draw[thick] (0,-2) circle (.075);
		
		\draw[thick] (-1.15,-0.02) -- +(0.6,-0.47);
		\draw[thick] (-1.15,-1) -- +(0.55,0);
		\draw[thick] (-1.15,-1.98) -- +(0.6,0.47);
		\draw[thick] (1.15,-0.02) -- +(-0.6,-0.47);
		\draw[thick] (1.15,-1) -- +(-0.55,0);
		\draw[thick] (1.15,-1.98) -- +(-0.6,0.47);
		\draw[thick] (0,-1.95) -- +(0.0,0.35);
		\draw[thick] (0,-0.05) -- +(0.0,-0.35);
		
		\node[text width=3cm, align=right] at (-3,-2.5) {\normalsize MCU\vphantom{j}};
		\node[text width=3cm, align=center] at (0,-2.5) {\normalsize Memory};
		\node[text width=3cm, align=left] at (3,-2.5) {\normalsize Components};
		
		\node[text width=3cm, align=right] at (-3,-1) {\normalsize Sensors};
		\node[text width=3cm, align=left] at (3,-1) {\normalsize Actuators};
		\node[text width=3.5cm, align=right] at (-3.25,0.5) {\normalsize Firmware/(RT)OS\vphantom{j}};
		\node[text width=3cm, align=center] at (0,.5) {\normalsize Application};
		\node[text width=3cm, align=left] at (3,.5) {\normalsize Web server\vphantom{j}};
		
		\draw (0.25,3) node[fill=\clic,minimum height=3cm, minimum width=9.4cm] (){};
		\node[draw, rotate=90, minimum width=2.95cm, minimum height=1.1cm, fill=\clict] at (-5,3) {};
		\node[text width=2.75cm, align=center, rotate=90] at (-5,2.9) {\normalsize Local \& Internal Communication};
		\node[text width=3cm, align=left] at (-3.9,4.25) {\textbf{\footnotesize Zone 2}};
		\node[text width=3cm, align=center] at (-3,3.75) {\normalsize USB};
		\node[text width=3cm, align=center] at (-1.5,3.25) {\normalsize JTAG};
		\node[text width=3cm, align=center] at (-3,2.75) {\normalsize RS232};
		\node[text width=3cm, align=center] at (-1.5,2.25) {\normalsize SPI};
		\interface{0}{3}	
		\node[text width=3cm, align=center] at (1.5,4) {\normalsize USB stick};
		\node[text width=3cm, align=center] at (3,3.5) {\normalsize SD card};
		\node[text width=3cm, align=center] at (1.5,3) {\normalsize Display};
		\node[text width=3cm, align=center] at (3,2.5) {\normalsize Microcontroller};
		\node[text width=3cm, align=center] at (1.5,2) {\normalsize PC};
		
		
		\draw (0.25,6) node[fill=\cpc,minimum height=3cm, minimum width=9.4cm] (){};
		\node[draw, rotate=90, minimum width=2.95cm, minimum height=1.1cm, fill=\cpct] at (-5,6) {\begin{minipage}{2.55cm}\begin{center}\normalsize Process \& Control\end{center}\end{minipage}};	
		\node[text width=3cm, align=left] at (-3.9,7.25) {\textbf{\footnotesize Zone 3}};
		\node[text width=3cm, align=center] at (-3,7.25) {\normalsize PROFINET};
		\node[text width=3cm, align=center] at (-1.5,6.75) {\normalsize EtherNet/IP};
		\node[text width=3cm, align=center] at (-3,6.25) {\normalsize Modbus};
		\node[text width=3cm, align=center] at (-1.5,5.75) {\normalsize CAN};
		\node[text width=3cm, align=center] at (-3,5.25) {\normalsize HART};
		\node[text width=3cm, align=center] at (-1.5,4.75) {\normalsize PROFIBUS};
		\interface{0}{6}
		\node[text width=3cm, align=center] at (1.5,6.75) {\normalsize PLC};
		\node[text width=3cm, align=center] at (3,6.25) {\normalsize Sensor};
		\node[text width=3cm, align=center] at (1.5,5.75) {\normalsize Actuator};
		\node[text width=3cm, align=center] at (3,5.25) {\normalsize HMI};
		
		
		\draw (0.25,9) node[fill=\cma,minimum height=3cm, minimum width=9.4cm] (){};
		\node[draw, rotate=90, minimum width=2.95cm, minimum height=1.1cm, fill=\cmat] at (-5,9) {\begin{minipage}{2.75cm}\begin{center}\normalsize Monitor \& Analyze\end{center}\end{minipage}};	
		\node[text width=3cm, align=left] at (-3.9,10.25) {\textbf{\footnotesize Zone 4}};
		\node[text width=3cm, align=center] at (-1.5,9.75) 
		{\normalsize WiFi};
		\node[text width=3cm, align=center] at (-3,9.25) 
		{\normalsize Ethernet};
		\node[text width=3cm, align=center] at (-1.5,8.75) 
		{\normalsize Bluetooth};
		\node[text width=3cm, align=center] at (-3,8.25) 
		{\normalsize 5G};
		\interface{0}{9}
		\node[text width=3cm, align=center] at (3,10) 
		{\normalsize Cloud};
		\node[text width=3cm, align=center] at (1.5,9.5) 
		{\normalsize SCADA};
		\node[text width=3cm, align=center] at (3,9) 
		{\normalsize Workstation};
		\node[text width=3cm, align=center] at (1.5,8.5) 
		{\normalsize Historian};
		\node[text width=3cm, align=center] at (3,8) 
		{\normalsize Smartphone};

		\end{tikzpicture}
		\caption{Different zones of a device and their respective interfaces with interaction systems.}
		\vspace*{-3mm}
		\label{fig:Interfaces}
	\end{figure}
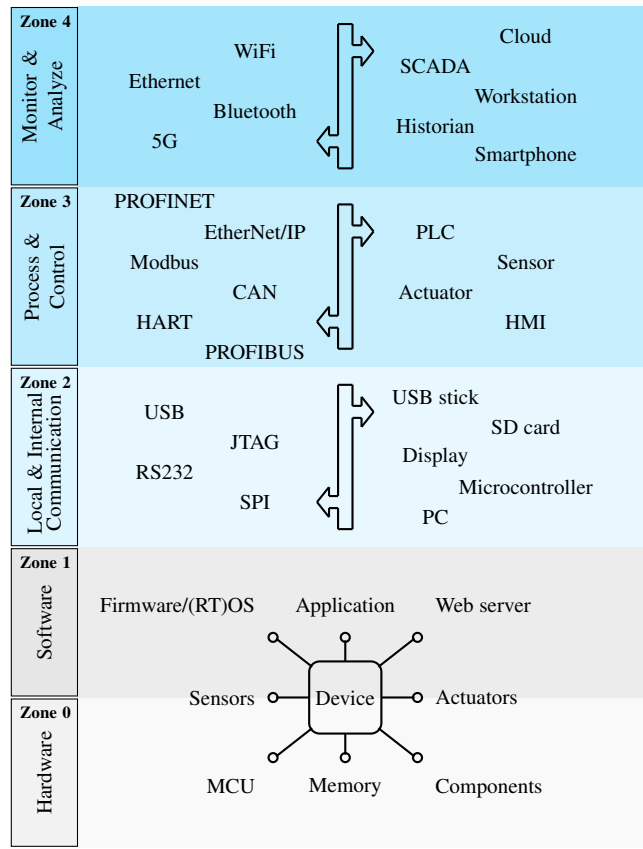

	IIoT devices are becoming increasingly complex. As a result of the IoT, new communication interfaces are being integrated that were previously rarely or never used in OT. In any case, they provide typically several interfaces for specific requirements. For better illustration, we have structured the various interfaces into zones in Figure~\ref{fig:Interfaces}. Zones~0 and 1 describe the hardware and software of a device. In zones 2 to 4, established communication protocols are listed in the left-hand column while systems that interact with them are listed in the right-hand column.
	
	In the following section, three possible attack vectors are introduced. Examples are used to illustrate how attackers from the different zones could proceed or how they could have an impact on other devices in these zones.
	
	\paragraph{Device attacks}
	
	In zone~0, device components can be physically manipulated. This may be intentional or accidental. In the latter case, a burnt-out circuit board or a defective engine could be replaced by a spare part that was not purchased from the original manufacturer for price reasons. Compatibility of hardware or software is not guaranteed for these components causing faulty operation, DoS and even destruction to result.
	
	As discussed in Section \ref{sec:Sources}, IIoT devices are especially threatened by highly capable actors. Attacks with high complexity and effort should consequently not be ignored. Costly invasive hardware attacks, such as probing, or rather cheaper non-invasive attacks, such as side-channel analysis, enable access to secret data, e.g., cryptographic keys. Attackers can also directly access the flash memory or EEPROM via interfaces from zone~2, e.g., JTAG. First, this allows them to read the memory to retrieve the firmware, i.e., intellectual property theft. Second, data or configurations can be modified, e.g., access data. Third, firmware can be exchanged so that arbitrary code can be executed. Attacks of this kind are complex, but they can cause considerable damage. In case the necessary knowledge is lacking, there are appropriate service providers for this (e.g., \href{www.break-ic.com}{www.break-ic.com}).
	
	The popular USB interface also enables multiple attacks. USB sticks can be used, for example, to load malware or destroy badly protected power and data lines, i.e., kill USB sticks. With bad usb devices, such as Hak5's rubber duckies, it is also possible to execute arbitrary commands and thus manipulate the device.
	
	\paragraph{Application attacks}
	
	The process of a production plant can be interrupted or stopped if the application of IIoT devices do not work properly. An attacker might change the configuration or move an actuator incorrectly via its display. The devices are often misconfigured as they still have the default or a trivial password. Many IIoT devices can also be programmed using a PC-based configuration tool. A common design flaw is that users must not be authorized to carry out these changes. As a result, it is often possible to reconfigure, update or reset a device by connecting to it via a cable or network. When such a vulnerability is exploited, it is difficult to reconstruct and verify the incident, as the devices often do not support user identification.
	
	Wrong commands can also originate from the devices of zone~3. The source can be either an already compromised PLC or a completely different device. Since messages of the most proprietary protocols are not authenticated, a different sender address can be spoofed. Reversely, incorrect information can also be sent to PLCs or HMIs. For example, PLCs from the manufacturer Schneider can be stopped using a simple command via the Modbus protocol \cite{Schneider}. The consequences of this abrupt stop may be catastrophic. Faulty commands or sensor data can also be sent to systems in zone 4, e.g., the SCADA system or the cloud. Since more decisions will be made by a data-driven Artificial Intelligence (AI) in the future, wrong choices may result.
	
	Due to the more widespread network protocols in zone~4, vulnerabilities can also be exploited remotely. Such vulnerabilities can be located in the firmware/operating system or the application. An example of the former are the Treck TCP/IP stack vulnerabilities called Ripple20 that allow remote code execution, which were recently discovered \cite{Ripple20}. Vulnerabilities in the application can be caused by a web server that allows SQL injection, for instance. Once they have successfully exploited a vulnerability, process operations can be sabotaged.
	
	\paragraph{Network attacks}
	
	The vulnerabilities just mentioned also allow an infection of botnets. If several devices in a network are infected and the botnet operator launches a DDoS attack, internal network traffic can be delayed. This can, for example, interrupt the connection of PLCs to the SCADA system. In case the attack is targeted at the own global company cloud, other plants might be affected as well.
	
	If the device is a network node, such as an edge device, this also results in multiple threats. Besides sniffing or tampering with messages, they can also be delayed or blocked. Especially for systems that have to meet real-time constraints, this can become a major threat.
	
	The network is also useful for spreading an infection. Especially the systems in zone~4 are targeted either for monetary gain through a ransomware attack or to obtain as much control as possible. Workstations with Win 7 or Win XP are not rare in ICSs, and thus this is often not much effort for an attacker.
	
	\section{Recommended Procedure}
	\label{sec:Procedure}
	
	Finally, we summarize all the previously discussed aspects to define a recommended procedure for the threat analysis.
	
	\quad
	\textit{1)  Know your device:} It is important to know the IIoT device in depth. Which operating system and third party libraries are utilized? Does it include actuators and sensors and/or is it collecting data from other devices (i.e., edge device)? How is the setup? What other equipment is connected to it? Is it connected to the Internet directly or through a gateway? Is it installed in critical infrastructures? What additional (PC-)tools are available for the device?
	
	\quad
	\textit{2) Creation of a network diagram:} A network diagram including all interfaces of the device can help identify which other systems it interacts with. The authorization should be specified for each entry and exit point, i.e., which actions can be performed and by whom. This is especially important for industrial protocols, such as PROFINET. While most IoT applications allow to implement security measures manually, it is not possible with these proprietary protocols. 
	
	\quad
	\textit{3) Identification and ranking of assets:} Which security goal is the most important one? Is the focus on maximum availability, authenticity of actions or privacy of user data? First, this is important to prioritize the exploration of vulnerabilities, and second, to subsequently find an appropriate mitigation measure. The latter is particularly relevant when safety must be guaranteed, as real-time behavior and encryption may not be feasible on a low-power IIoT device.
	
	\quad
	\textit{4) Identification of threat sources:} Who is interested in attacking the device and what are their motives? This is useful for deliberately including or excluding types of attacks. For IIoT devices in critical infrastructures, the more complex invasive and non-invasive hardware attacks should be addressed.
	
	\quad
	\textit{5) Identification of threats and vulnerabilities:} The next step is to identify threats and vulnerabilities. Table~\ref{tab:ThreatsVuls} serves as a kick-off aid. In general, we can consider attacks on identification and authentication, authorization, availability, system, data and communication integrity, data confidentiality, privilege escalation and repudiation. Penetration testing can be used to discover additional vulnerabilities, but also to verify those already identified and show their severity.
	
	Using attack scenarios, attacks can be better reconstructed in retrospect. For example, the threat \textit{setting an invalid communication configuration} results in a \textit{denial of service}. The attack vector is that the \textit{web server} is accessible via the \textit{Ethernet} interface. The action \textit{changing of communication parameter} has the consequence that the \textit{connection to PLCs is terminated}. The utilized vulnerability is a \textit{default password} that results in a \textit{privilege escalation}. Additional notes, such as \textit{default password can be found in the manual}, can also be useful.
	
	\quad
	\textit{6) Vulnerability and risk assessment:} To rate a vulnerability, all threats and their consequences from the different attack scenarios should be considered. Using the Common Vulnerability Scoring System (CVSS), the severity of vulnerabilities can be expressed by a number. For risk assessment, it is advisable to consider not only the severity of the vulnerability but also its likelihood and impact.
	
	\section{Conclusion and Further Work}
	\label{sec:Conclusion}
	
	Compared to IoT equipment, IIoT devices are at increased risk, since they are part of the OT that controls physical processes. Beside high availability, safety is also particularly important in these applications. In addition to destroying a production facility, people can be injured and a population can even be cut off from the power grid.
	
	Several threat sources and their motives were presented and ranked using examples. It turned out that the most serious threat originates from government-sponsored actors, who often target critical infrastructures. Afterwards, numerous threats and vulnerabilities were listed, which exist among other reasons, because security was ignored in the industrial sector for decades. Among the threats, destruction caused by moving parts and intellectual property theft must be highlighted, while the vulnerabilities include manipulation of the hardware and the frequently insecure communication. Lastly, we provided a procedure for identifying and assessing threats and vulnerabilities that emphasizes the specialties of IIoT devices. In order to prevent these, we intend to develop countermeasures for low-power IIoT devices as the next step.
	
	\section*{Acknowledgment}
	
	The research project ``Intelligent Security for Electrical Actuators and Converters in Critical Infrastructures (iSEC)'' is a collaboration of SIPOS Aktorik GmbH, Grass Power Electronics GmbH and OTH Amberg-Weiden. It is supported and funded by the Bavarian Ministry of Economic Affairs, Regional Development and Energy.
	
	\balance
	
\end{document}